\documentclass[12pt,preprint]{aastex}
\newcommand{\myemail}{ruta@rri.res.in}
\slugcomment{Accepted in ApJ}
\shorttitle{Spectral index images of A2256}
\shortauthors{Kale \& Dwarakanath}
\begin{document}
\title{Spectral Index Studies of the Diffuse Radio Emission in \\ Abell 2256: Implications to Merger Activity}
\author{Ruta Kale and K. S. Dwarakanath}
\affil{Raman Research Institute, Bangalore 560080, India}
\email{\myemail, dwaraka@rri.res.in}
\begin{abstract}
We present a multi-wavelength analysis of the merging rich cluster of galaxies Abell 2256. We have observed A2256 at 150 MHz using the Giant Metrewave Radio Telescope and successfully detected the diffuse radio halo and the relic emission over an extent $\sim1.2$ Mpc$^2$.
Using this 150 MHz image and the images made using archival observations from the VLA (1369 MHz) and the  WSRT (350 MHz), we have produced spectral index images of the diffuse radio emission in A2256. These spectral index images show a distribution of flat spectral index (S$\propto\nu^\alpha$, $\alpha$ in the range -0.7 to -0.9) plasma in the NW of the cluster centre. Regions showing steep spectral indices ($\alpha$ in the range -1.0 to -2.3) are toward the SE of the cluster centre. These spectral indices indicate synchrotron life times for the relativistic plasmas in the range 0.08 - 0.4 Gyr. We interpret this spectral behaviour as resulting from a merger event along the direction SE to NW within the last 0.5 Gyr or so. A shock may be responsible for the NW relic in A2256 and the Mpc scale radio halo towards the SE is likely to be generated by the turbulence injected by mergers. 
Furthermore, the diffuse radio emission shows spectral steepening toward lower frequencies. This low frequency spectral steepening is consistent with a combination of spectra from two populations of relativistic electrons created at two epochs (two mergers) within the last $\sim$0.5 Gyr. Earlier interpretations of the X-ray and the optical data also suggested that there were two mergers in Abell 2256 in the last 0.5 Gyr, consistent with the current findings. 
Also highlighted in this study is the futility of correlating the average temperatures of thermal gas and the average spectral indices of diffuse radio emission in respective clusters. 
\end{abstract}
\section{Introduction}
The intra-cluster medium (ICM), which pervades the space between the galaxies in galaxy clusters is 
known to consist of hot thermal gas ($\sim 10^8$ K), magnetic fields ($\sim 1\mu$G) and relativistic particles. The thermal gas emits mainly in X-rays by thermal Bremsstrahlung mechanism. 
A direct evidence for magnetic fields and relativistic electrons in the ICM are the extended diffuse synchrotron sources associated with the ICM detected in a fraction of clusters (Ferrari et al 2008).
These are classified broadly as radio halos (centrally located in clusters and unpolarized ($\lesssim 5\%$)) and radio relics (filamentary/ arc-like, located at cluster peripheries and polarized ($\sim20-30\%$)) (Ferrari et al 2008). So far, such sources have been detected in merging clusters of galaxies; thus a strong connection between their origin and merger is favoured (Ferrari et al 2008). The primary models predict that shocks and/or turbulence induced during merging events reaccelerate the electrons in the ICM: shocks reaccelerate by Fermi I process (Ensslin et al 1998) and/or 
adiabatic compression of fossil radio plasma (Ensslin \& Gopal-Krishna 2001); turbulence reaccelerate 
via stochastic, Fermi II or MHD waves (Brunetti et al 2004; Cassano \& Brunetti 2005). The secondary models regard the synchrotron emission to come from electrons generated in hadronic collisions (Dolag \& Ensslin 2000 and references therein); this model awaits observational evidence and is not discussed here. 
\par The shocks and/or turbulence in the ICM which are responsible for accelerating charged particles leave signatures in the spectra of accelerated particles. A shock can leave a trail of accelerated charged particles in its wake; turbulence can lead to patchy distribution of spectral index in the source (Feretti et al 2004 and references therein). Synchrotron spectral index steepens with time as a result of energy losses proportional to E$^2$. 
 In a simplistic approach, the spectral index can be used to estimate the time since the relativistic plasma was accelerated-- the spectral age. If the break frequency of the spectrum can be identified, the spectral age can be estimated; if it cannot be identified, at least a limit on the spectral age can be obtained. Spectral age distribution across the extents of radio halos and relics can then be connected to the geometry of shock passage or the sites of efficient/inefficient turbulent acceleration. Due to the complex nature of the reacceleration by turbulence, this may not be the best estimate of the time since the acceleration took place.
 Consistency with the merger geometries proposed based on X-ray surface brightness and optical galaxy distributions can be verified.
Such studies require good quality multi-frequency maps of radio halos and relics. Radio halos and relics being extended ($\sim5$ to $30$ arcminutes for clusters in redshift range $0.2 - 0.02$), low surface brightness ($\sim$ 1 mJy arcmin$^{-2}$ at 1.4 GHz) sources are difficult to image at multiple frequencies with comparable sensitivities. Spectral index maps have been constructed only in a few clusters so far (for example, Coma (Giovannini et al 1993), A665, A2163 (Feretti et al 2004), A3562 (Giacintucci et al 2005), A2744 and A2219 (Orru et al 2007) and A2255 (Pizzo \& Bruyn 2009)). These have shown features like radial spectral steepening and patchy distribution of spectral indices and have been interpreted as variations in the magnetic field in the cluster and turbulence by the authors. Various geometries of merger have also been discussed.
 It is important to realise that the synchrotron spectrum is curved by nature (energy losses proportional to E$^2$). To understand the curvature in the spectra of different regions and identification of the break frequencies, spectral index maps between multiple frequencies are necessary. So far a study with two spectral index maps has been carried out for the cluster A2255 by Pizzo \& Bruyn (2009). Radio halos and relics are rare sources and thus only a few clusters where such detailed study can be carried out are available. \par One such cluster is Abell 2256 (hereafter A2256) which is a host to a radio halo and a radio relic and shows clear signatures of merger.
 Using two spectral index maps (150-350 MHz and 350-1369 MHz), a study of the complex dynamics in the cluster A2556 is presented here.
  A2256 is a rich, X-ray luminous ($L_{X [0.1 - 2.4 keV]}\sim3.8\times10^{44}$ erg s$^{-1}$, Ebeling et al 1996) galaxy cluster at a redshift of 0.0581 (Struble \& Rood 1999). The X-ray surface brightness is elongated in the east-west direction and shows substructures (Briel et al 1991; Sun et al 2002); the radial velocity distribution of galaxies shows the 
presence of three distinct groups of galaxies (Berrington et al 2002). These have been interpreted to be indicators of ongoing merger in A2256. Recent temperature maps of A2256 with $Chandra$ (Sun et al 2002) and XMM Newton (Bourdin \& Mazzotta 2008) show variation between 4 - 10 keV in $\sim0.6$ Mpc region around the cluster centre. 
 Apart from several head-tail radio galaxies, A2256 hosts diffuse radio emission in the north-west of the cluster centre (the radio relic) and at the centre (the radio halo). It has been studied in radio wavelengths for the past three decades (Bridle \& Fomalont 1976; Bridle et al 1979; Rottgering et al 1994 (hereafter R94); Clarke \& Ensslin 2006 (hereafter CE06); Brentjens 2008 (hereafter B08)). Polarization of $\sim20-40\%$ was detected at 1400 MHz in the relic by CE06; at 350 MHz it is unpolarized ($<1\%$) (B08).  
A2256 also hosts the peculiar steep spectrum source 'F'; the optical identification of it is still being debated (B08). New steep spectrum sources have been detected in the 330 MHz images of A2256 made using the GMRT by van Weeren et al (2009); these are at the periphery of the cluster and are unrelated to the halo and the relic that will be discussed in this paper. 
\par Study of the integrated spectrum (including the radio galaxies, the compact radio sources as well as the diffuse radio emission) of A2256 has shown that it steepens at low frequencies (B08); a property unique to this cluster. The properties of the relic in A2256 are inconsistent with the scenario of acceleration in structure formation accretion shocks 
(Ensslin et al 1998). They propose a shock radius of $\sim1$ Mpc for the geometry of the relic in A2256, which implies an origin in shocks interior to the cluster -- the merger shocks. Based on the simulations of Roettiger et al (1995), merger between clusters having mass ratios 2:1, in the direction northwest to southeast, such that the smaller cluster is moving towards the observer has been discussed in the case of A2256. This model was proposed to explain the X-ray properties and temperature distribution in A2256 as estimated by {\it ROSAT} (Briel \& Henry 1994); further sensitive X-ray measurements ({\it ASCA, Chandra, XMM Newton}) do not confirm the measurements of temperature by the {\it ROSAT} (Sun et al 2002; Bourdin \& Mazzotta 2008); rendering the model unusable. Based on X-ray substructure and identification of three distinct groups of galaxies in A2256, two mergers have been discussed (Sun et al 2002; Berrington et al 2002; Miller et al 2003). Berrington et al (2002) propose two mergers -- one between two comparable mass subclusters and another between a group and the primary cluster. Miller et al (2003) have favoured the possibility of the group being responsible for the radio relic whereas CE06 favour the merger of two subclusters. 
The case of A2256 is complex.  The spectral distribution across the diffuse radio emission in A2256 is presented in this paper and its implications to the mergers are discussed.
\par This paper is organized as follows. In Section 2, radio observations of A2256 using the Giant Metrewave Radio Telescope (GMRT) and the data reduction are described. The radio image at 150 MHz and the spectral index maps between 150-350 MHz and 350-1369 MHz are presented in Section 3. Discussion of these results in the contexts of temperature of the ICM and the dynamics in the cluster are presented in Section 4. The conclusions are presented in Section 5.	
\par We adopt a $\Lambda$-dominated cosmology with a Hubble constant, $H_0$, of 73 km s$^{-1}$ Mpc$^{-1}$,  $\Omega_M=0.27$ and $\Omega_{\Lambda}=0.73 $  which implies a distance scale of 
$\sim 70$ kpc arcmin$^{-1}$ at the redshift of A2256. 
\section{Radio Observations and Data Reduction}
The observations of A2256 were carried out with the GMRT (Oct '07) at 150 MHz for a duration of $\sim 7 $ hr with a bandwidth of 6 MHz. The Astronomical Image Processing System (AIPS) was used to analyse this data.
The GMRT data at 150 MHz are affected by radio frequency interference (RFI). Using the data visualization and editing tasks in AIPS, data affected by RFI were identified and removed. About $35\%$ of the data were excised. Standard procedures of absolute flux density, gain and bandpass calibrations were carried out. This calibrated data was imaged and several self-calibration iterations were performed to obtain the best images. Use of uniform and natural weighting of the visibilities resulted in images at 150 MHz at resolutions of $20''\times20''$ and $33''\times33''$ with rms sensitivities of $\sim2.2$ and $\sim2.5$ mJy beam$^{-1}$, respectively. 
\par Data in the form of visibilities at 1400 MHz (project code AC522) were obtained from the archives of the Very Large Array (VLA). This data contained observations for a duration of 6 hr with the VLA in the D configuration. A  bandwidth of 25 MHz at each of the 2 IFs, namely, 1369 and 1417 MHz, was used in these observations. Images at each of 1369 and 1417 MHz are published in CE06 (Fig. 1, top 2 panels). CE06 have detected the largest extent of the diffuse emission in the image at 1369 MHz and thus this frequency was chosen to obtain an image for our purpose. Visibilities at 1369 MHz were edited, calibrated and imaged using AIPS. Natural weighting of the visibilities resulted in an image at 1369 MHz with a resolution of $67''\times67''$ and an rms of 0.07 mJy beam$^{-1}$. 
This image is similar in terms of rms and detection of extended sources to that in Fig. 1 in CE06 (rms $\sim0.06$ mJy beam$^{-1}$; beam $\sim 52''\times45''$). 
\par Visibilities recorded in frequency bands around 350 MHz containing observations of A2256 were obtained from the Westerbrok Synthesis Radio Telescope (WSRT) archives.
This data contained $\sim11 $ hr observations of A2256 with the WSRT in a configuration with the shortest baseline of $\sim 72$ m. These visibilities were recorded in 8 frequency bands, between 310-390 MHz, each having a bandwidth of 10 MHz. Natural weighting of the visibilities resulted in a synthesized beam of $\sim62''\times62''$. For easy comparison with the 1369 MHz image, a beam with FWHM of $67''$ was chosen for imaging and an rms of 0.6 mJy beam$^{-1}$ was achieved using visibilities in all the frequency bands. This image is similar in quality to that in Fig. 2 in B08 produced from the same dataset. For spectral comparison, an image with a beam of $67''\times67''$ was produced at 150 MHz from the GMRT data. The flux density calibration errors are $\sim15\%$ at 150 MHz and $<10\%$ at 350 and 1369 MHz.
\par Primary beam corrections to images at each of the frequencies were applied using the task 'PBCOR' in AIPS. At 150 MHz and 1369 MHz, the coefficients of the polynomials representing the respective primary beams reported on the GMRT and the VLA websites were used. The primary beam of the WSRT is approximated by the  function $cos^6(\theta)$, where $\theta$ is a function of the frequency and the angular distance from the pointing centre. 
Using this function, parameters for the task 'PBCOR' appropriate for the WSRT were calculated and used. These primary beam corrected images were used for obtaining spectral index images between these frequencies.
\section{Results}
\subsection{Radio Images}
The central portion of the GMRT 150 MHz image of the A2256 region at a resolution of $67''\times67''$ is presented in Fig. 1. This image shows sources with extents ranging between $\sim 67''$ and $\sim 15'$. The sources are labelled according to the convention in Bridle et al (1979) and R94 for easy reference. The sources A, B, C, D and F have extents ranging between $\sim 67''$ and $\sim 6'$. The radio sources A, B, C and D are radio galaxies with optical counterparts in A2256 (R94; Miller et al 2003). 
 The source F is an extended ultra-steep spectrum source (Masson \& Mayer 1978; Bridle et al. 1979; R94). The extent of source F at 150 MHz is $\sim4'$ which is $\sim280 $kpc if it is at the redshift of A2256. The identification of source F as a radio tail of optically identified galaxy 122 of Fabricant et al (1989) is still being debated (B08).
 \par The sources G and H are extended ($\sim10'$) sources and are together termed as the ``radio relic'' in A2256 (R94). High resolution ($\sim 1.4''\times1.2''$) 20 cm maps with the VLA in the A configuration by R94 resolve out G and H completely; thus confirming their diffuse nature. No unique optical counterpart can be associated with these sources (Miller et al 2003). 
The radio relic seen at 150 MHz, covers a region of $\sim 1\times0.4$ Mpc$^2$. 
CE06 have reported an extent of $\sim 1.125\times 0.52$ Mpc$^2$ for the relic. 
The larger extent reported by CE06 and also detected in our 1369 MHz image covers a region of extent $\sim100$ kpc toward the northwest and north of the boundaries of the relic shown in Fig. 1.
The surface brightness of this $\sim100$ kpc region is $\sim 0.4$ mJy arcmin$^{-2}$ at 1400 MHz (CE06). This surface brightness implies a surface brightness of 2.4 mJy arcmin$^{-2}$ at 150 MHz assuming a spectral index of -0.8. This is $\sim$6 times lower than the sensitivity in the 150 MHz image presented here. The diffuse emission pervading the central region around source D is the ``radio halo`` (CE06). The radio halo and the relic emission overlap and thus it is difficult to determine the exact extents of each. 
The region towards south-east that is detected at 1369 MHz by CE06 ($\sim$ 0.2 mJy arcmin$^{-2}$) and at 350 MHz by B08 is beyond the SE boundary shown in Fig. 1 and is not detected at 150 MHz due to its low surface brightness ($\sim$1.2 mJy arcmin$^{-2}$ at 150 MHz). 
\subsection{Integrated spectrum}
\par The total flux densities inclusive of the radio halo, the radio relic and the discrete sources are $10.0\pm1.5$, $3.6\pm0.1$ and $1.47\pm0.07$ Jy at 150, 350 and 1369 MHz, respectively. The total flux densities were measured using the same area at all the three frequencies.
In Table 2 of B08, total flux density measurements of A2256 at frequencies ranging from 22.25 to 2695 MHz have been reported. Our estimates at 350 and 1369 MHz are within $\sim1\sigma$ errors of the values reported by B08.
The total flux density recovered at 150 MHz is $\sim25\%$ higher than that reported by B08 ($8.1\pm0.8$ Jy at 151 MHz from Masson \& Mayer (1978)).  
These total flux density estimates imply spectral indices of $\alpha^{350}_{150}=-1.20\pm0.13$ and $\alpha^{1369}_{350}=-0.65\pm0.01$ for the integrated spectrum of A2256. Total flux densities of $17\pm2$ Jy at 81.5 MHz (Branson 1967) and $3.51\pm0.06$ Jy at 351 MHz (B08) imply a spectral index of  $\alpha_{81.5}^{351}=-1.08\pm0.06$. Our estimate of spectral index between 150 and 350 MHz is consistent within $1\sigma$ error with this low frequency spectral index. 
The total flux density of A2256 at 1369 MHz is not available in CE06. Using the total flux density estimate at 351 MHz from B08 and our estimate at 1369 MHz, the spectral index of A2256 is $\alpha^{1369}_{351}=-0.63\pm0.01$; our estimate ($\alpha_{350}^{1369}$) is consistent within $1\sigma$ error. 
\subsection{Spectral index maps}
\par The (u,v)- coverages at 150, 330 and 1369 MHz were comparable. The (u,v)-plane was sampled sufficiently to short baselines $\sim$0.1 k$\lambda$ to image extents of $\sim20'$ at the three frequencies. With the use of the VLA-D configuration at 1369 MHz, the WSRT at 350 MHz and the GMRT at 150 MHz, such a (u,v)-coverage could be achieved. Total flux densities were recovered at all the frequencies and thus these images were considered suitable for producing spectral index maps.
The spectral index maps of A2256 were obtained using the primary beam gain corrected images produced at 150, 350 and 1369 MHz having resolutions of $67''\times67''$. Pixels having flux densities less than $5\sigma$ level in the respective images were blanked before making spectral index maps to minimise uncertainties. Thus, pixels having flux densities less than 18 mJy beam$^{-1}$ at 150 MHz, 3 mJy beam$^{-1}$ at 350 MHz and 0.35 mJy beam$^{-1}$ at 1369 MHz were blanked. To obtain the spectral index map of the radio halo and the relic, subtraction of the radio sources (A, B, C, D and F) contaminating the diffuse emission was attempted. Visibilities of short baselines were excluded to obtain images of only the discrete sources.  
Such images were subtracted from the respective images made using all the visibilities at each of the frequencies. 
This procedure resulted in either a partial subtraction of the relic or an incorrect subtraction of the discrete sources.
 As an alternative, models including upto 4 Gaussian components were used to fit the discrete sources. Those were subtracted from the images. This subtraction also resulted in artefacts in the images. The discrete sources have complex structures and require sophisticated modeling. To minimise the uncertainties, no discrete sources were subtracted from the final maps that were used in making the spectral index maps. The positions of discrete sources will be kept in mind while discussing the spectra of diffuse sources.
\par We present the spectral index maps of A2256 in Fig. 2 and the corresponding error maps in Fig. 3. The extent of the spectral index map between 150 and 350 MHz (Fig. 2, right) is limited by the emission detected at 150 MHz. As can be seen in Fig. 3, the typical error in each of  the spectral index maps is $<0.10$. The error is uniform except at the edges of the radio emission where it is between $0.2$ and $0.3$ in the 350-1369 MHz spectral index map and between $0.2$ and $0.5$ in the 150-350 MHz spectral index map. The contours plotted in Fig. 3 are of the images of A2256 at 350 MHz (left panel) and at 1369 MHz (right panel) obtained using the archival data as described in Sec. 2. See Sec. 2 for the comparison of these with the images published by CE06 and B08. These images were used in making the spectral index map presented in Fig. 2 (left).
\par  Before studying the spectra of the diffuse radio emission, we examined the spectral behaviour of the discrete sources for consistency with earlier measurements. The discrete sources, except F, have spectral indices $\sim -0.7\pm0.1$ which are typical of radio galaxies and can be seen in Fig. 2. 
The components of the source F, namely F1, F2 and F3, discussed in detail by B08 are marked in the spectral index maps (Fig. 2) . The source F2 has spectral indices of $\alpha_{350}^{1369}=-1.80\pm0.05$ and $\alpha^{350}_{150}=-1.10\pm0.05$. These values of spectral indices of F2 are consistent with the values 
 $\alpha^{1446}_{610}=-1.71\pm0.08$ and $\alpha^{350}_{150}=-1.20\pm0.05$ obtained by B08.
The tail of source C extends in the north-west direction and is visible in the spectral index map (Fig. 2, left) as a linear feature. The spectral index, $\alpha^{1369}_{350}$, gradually steepens from $\sim -0.8\pm0.1$ near the head (marked C) to $\sim -1.20\pm0.05$ in the tail. Such a spectral steepening from the head towards the lobes has been seen in radio galaxies and is believed to be due to spectral ageing. Bridle et al (1979) term the tail of C as source ``C(ii)'' and report a spectral index of $-1.04\pm0.13$ between 610 and 1415 MHz. Our estimate is consistent with that of Bridle et al (1979). Note that the discrete source O has a spectral index $\sim -0.7$ over the range of 150 to 1369 MHz (Fig. 2) as expected. Source L is not detected at 150 MHz. 
While discussing the spectral indices of diffuse emission, the regions having large errors ($\sim0.4-0.5$, see Fig. 3 for error maps) are not considered.
A complex distribution of spectral indices is seen over the extent of the diffuse radio emission (Fig. 2). Occurrence of flat spectral indices ($\alpha^{1369}_{350}\sim-0.7$ to $-0.9$ and $\alpha^{350}_{150}\sim-0.7$ to $-1.1$)
is noticed in the northwest (NW) regions marked G and H of the diffuse emission. Towards southeast (SE) of the discrete source D the spectra are steeper ($\alpha^{1369}_{350}\sim-1.2$ to $-2.3$ and $\alpha^{350}_{150}\sim-1.5$ to $-2.5$).
To establish the significance of this trend, 
 six slices across each of the spectral index maps in the direction NW-SE (approximately $30^\circ$ clockwise from the north) separated by $2'$ from each other in the perpendicular direction were taken. Plots of spectral index versus distance from the SE edge of the diffuse radio emission were produced. 
Towards the NW, the diffuse radio emission shows average spectral indices of $\alpha^{1369}_{350}\sim-0.8\pm0.05$ and $\alpha^{350}_{150}\sim-0.9\pm0.2$.  
Toward SE of the source D, the average spectral indices are $\alpha^{1369}_{350}\sim-1.4\pm0.1$ and $\alpha^{350}_{150}\sim-2.3\pm0.2$. 
\par 
As noted in earlier works the integrated spectrum of A2256 which included the diffuse as well as discrete sources showed steepening at lower frequencies (B08). To find out whether such a steepening occurs in the  diffuse radio emission, independent regions, each having a size of the synthesised beam, were chosen along the SE-NW direction (broken line in Fig. 2, right). The regions affected by the discrete sources were avoided. Flux densities of the chosen regions at each of 150, 350 and 1369 MHz were estimated from the images and plotted. These spectra are presented in Fig. 4. The flux densities have been scaled and shifted along the y-axis such that the topmost spectrum is of the region extreme NW and the spectrum at the bottom is of the region extreme SE. 
It was found that each of these independent patches of diffuse radio emission have steeper spectral indices between 150 and 350 MHz as compared to that between 350 and 1369 MHz; except for the extreme NW where the spectrum is straight. From the NW to the SE, $\alpha_{150}^{350}$ steepens from $-0.60$ to $-2.30$. The value of $\alpha_{350}^{1369}$ shows mild variation in the first 3 curves from the top (Fig. 4) but steepens to $-1.23$ further towards SE edge. It was also noted that the difference, $|\alpha_{150}^{350} -\alpha_{350}^{1369}|$, increases from NW to SE (Fig. 4).
\section{Discussion}
Most galaxy clusters with radio halos and relics also show signatures of recent or ongoing merger activities (Ferrari et al 2008). The cluster A2256 is a complex case involving more than one merger (Berrington et al 2002; Sun et al 2002).
The implications to the spectral index trends in the diffuse radio emission in A2256 from the proposed origins in merger shocks and/or in turbulence are explored further here. 
The diffuse radio emission in A2256 has been referred to as the relic (G \& H in Fig. 1) and the halo (the region S and SE of the source D in Fig. 1) in earlier works. 
\subsection{Spectral Index and ICM Temperature}
\par A schematic representation of the temperature map of A2256 (Bourdin \& Mazzotta 2008) is shown by the black (white in online version) contours in Fig. 2 (left). In A2256, the spectral index of the diffuse emission varies over the range -0.7 to -2.5 and the temperature of X-ray emitting gas varies over the range 4 - 10 keV. It is found that the steep spectrum ($<-1.5$) region towards the SE is co-spatial with the hottest ($\sim10$ keV) region in the cluster. The NW region having a flat spectrum ($\sim-0.8$) has temperatures $\sim4-7$ keV. The cluster A2256 does not show any correlation between the hot X-ray and the flat radio spectrum regions; in fact the SE region is a clear anti-correlation. 
Cooling times of thermal gas in merging clusters are typically more than the lifetime of the cluster (Hubble time) (Buote 2002).
Synchrotron cooling times are at least 10-100 times shorter than a Gyr for break frequencies in the GHz range and typical intra-cluster magnetic fields. The thermal plasma will essentially be at the same temperature while the synchrotron spectrum ages (steepens) and finally fades.  
\par Earlier studies have explored possible correlations between the temperature of the thermal gas and the  spectral indices of radio halos confined in them. Feretti et al (2004) report the absence of one to one correspondence between the high temperature and the flat ($\sim-0.8$) synchrotron spectrum regions in the radio halo in A665 and only a mild correspondence in A2163. This can be easily understood by comparing the cooling time of the thermal plasma and that of the synchrotron plasma. 
Recently,  Giovannini et al (2009) have reported a mild correlation between the average temperatures of the ICM and the average spectral indices of the radio halos in the respective clusters. 
As seen in A2256, the temperature varies over a range of 4 - 10 keV and the spectral index varies over a range of $-0.7$ to $-2.5$. Similar variations have been found in other clusters (Feretti et al 2004; Govoni et al 2002; Bourdin \& Mazzotta 2008) too. Therefore comparing the average values of ICM temperatures and of spectral indices of radio halos in respective clusters can be misleading. Moreover, the estimates of temperatures in clusters and the co-spatial occurrence of radio emission and hot gas are affected by projection effects. 
\subsection{Spectral index and cluster dynamics} 
 The implications of the properties of the complex spectral index distribution in A2256 to the geometries and timescales of mergers are discussed here. The diffuse radio emission in A2256 shows the presence of two regions in spectral index maps. A region NW of the discrete sources A, B and D with flat spectral indices and another SE of these sources having steep spectral indices (Fig. 2). 
In order to estimate spectral age of radio emission a knowledge of the magnetic field in that region and of the break frequency is required. In A2256, the estimates of magnetic field based on the depolarization properties of the filament G in the NW are in the range 0.02 - 2 $\mu$G (B08) and those of the SE region based on the classical minimum energy, equipartition and hadronic minimum energy conditions range between $1.5 -8 \mu$G 
(CE06). 
For simplicity, a typical value of 1$\mu$G for cluster magnetic field (Carilli \& Taylor 2002) is used. Synchrotron spectrum is curved and requires measurements at several frequencies, over the entire range from a few MHz to tens of GHz, to identify the break frequency.
Flux density estimates of regions within the diffuse radio emission in A2256 are available from the images at 150, 350 and 1369 MHz as presented in this paper; the break frequencies cannot be identified using these. 
However, from the spectral indices, upper and lower limits on the break frequencies of different regions in the diffuse radio emission can be estimated. The values of spectral indices in the NW and the SE regions discussed in Sec. 3.3 imply that 1369 and 150 MHz can be considered as the lower and the upper limits on the break frequencies of these regions, respectively.
Using these break frequencies, the upper and lower limits on the spectral ages of the NW and of the SE regions of the diffuse radio emission 
are $\sim$ 0.08 and 0.4 Gyr, respectively. 
The spectral ages imply that the radio emission in the NW relic region is young and the acceleration is very efficient. 
The relic could be the present location of a shock front. One possibility is that a cluster merger in the direction SE to NW drove shocks and injected turbulence in the ICM along its way. Cluster merger shocks can accelerate particles to relativistic energies (Ensslin et al 1998; Hoeft \& Bruggen 2007). But the distances to which the shock accelerated particles can diffuse within their radiative lifetimes are short ($\sim200$ kpc, see Brunetti et al 2008). Thus shock acceleration alone cannot be responsible for the Mpc scale radio emission as seen in A2256.
However, the relic region having a sharp edge at the NW and a projected width of $\sim300$ kpc towards the SE could be the result of shock acceleration. 
\par The reasons for the possibility of shock acceleration in the relic region are as follows. An average linear polarization fraction of $\sim20\%$ has been detected across the relic at 1.4 GHz with the averaged B-vector oriented with an angle 10$^\circ$ (measured east of north) (CE06). In the region of shock, the magnetic fields become alligned with the shock plane. Based on the polarized fraction in the relic, CE06 propose that the shock plane is alligned at an angle of $\sim45^\circ$ with the plane of the sky. This orientation is consistent with the SE to NW direction in which the shock is likely to have propagated based on the spectral ages discussed above. 
At the location of the shock, which is the site of reacceleration, flat spectral indices are expected and gradually in the wake of the shock, the spectra would be steeper. The relic has a flat spectral index of $\alpha\sim -0.7$ between 150 and 1363 MHz but shows a gradual  steepening from NW to SE in the frequency range 1363 -1700 MHz (CE06).
 \par It should be noted that there is no evidence for a shock in A2256 from the X-ray observations with Chandra and XMM Newton (Sun et al 2002; Bourdin \& Mazzotta 2008). Viewing angle can be a reason for the non-detection of shock in X-rays. Nevertheless there is evidence for mergers in A2256. The distributions of the optical galaxies and of the X-ray emission in A2256 show substructures which are believed to be due to 2 merger events (Berrington et al 2003; Sun et al 2002). Based on the optical substructure, Berrington et al (2003) propose two mergers in A2256. One is a merger between the primary cluster (PC) and the sub-cluster (SC) and another between a group (Gr) and the combined system of PC and SC or the PC. 
 The symbols box, circle and star (Fig. 2, left) indicate the positions of the optical centroids of the PC, the SC and the Gr, respectively. The estimates of mass for the PC, the SC and the Gr are 1.6$\times10^{15}$M$_{\odot}$, 0.51$\times10^{15}$M$_{\odot}$ and 0.17 $\times10^{15}$M$_{\odot}$, respectively. Of the two mergers one is  major merger (PC+SC, mass ratio $\sim3$) and another is a minor merger (PC+Gr or (PC+SC)+Gr, mass ratio $\sim10$). Numerical simulations have shown that mergers with mass ratios $\sim3-10$ can give rise to shocks of Mach numbers $\sim1.5 - 3$, but these are not sufficient for accelerating particles that are responsible for observed synchrotron emission (Gabici \& Blasi 2003). Thus, even in the region of the relic in A2256, mechanisms other than shock acceleration may be active.
\par  
Turbulent reacceleration is a mechanism by which Mpc scale radio emission can be generated in clusters of galaxies (Roettiger et al 1999; Fujita et al 2003; Brunetti et al 2004; Brunetti \& Lazarian 2007). The first step for this mechanism is injection of fluid turbulence in the ICM. Mergers of clusters are one of the most favoured routes for injection of fluid turbulence on scales of 0.5-1 Mpc in the ICM.
Numerical simulations and observations have provided evidence for the existence of fluid turbulence in the ICM on the scales of 0.1 - 1 Mpc (Sunyaev et al 2003; Schuecker et al 2004; Churazov et al 2004; Gastaldello \& Molendi 2004; Dolag et al 2005; Vazza et al 2006).
 According to the simulations of cluster mergers by Cassano \& Brunetti (2005), mergers with mass ratios in the range 3-10, as is the case of A2256, inject energy equivalent to $5-8\%$ of that of thermal gas in large-scale fluid turbulence. 
\par The efficiency of turbulent reacceleration depends on many factors such as the timescale for cascading and wave-particle coupling which depend on many physical quantities which are unknown. For example, the spectrum of the magnetosonic (MS) waves and the structure of the magnetic fields are not known. However, Cassano \& Brunetti (2005) have shown that MS waves with scales $\sim 100$ kpc can efficiently accelerate fast electrons in the ICM to energies sufficient for producing the synchrotron emission detected in radio bands. The decay time of the MHD turbulence at injection length scales ($L_{inj}\sim$ 1Mpc) can be estimated using:
\begin{equation}
 \tau_{kk}(Gyr)\sim (\frac{v_i}{2\times10^3 
 kms^{-1}})^{-1} (\frac{L_{inj}}{1Mpc})(\frac{\eta_t}{0.25})^{-1}
\end{equation}
where $v_i$ is the relative velocity of impact of merging clusters and $\eta_t$ is the fraction of the energy in turbulence that is in MS waves (Cassano \& Brunetti 2005). The value of $\eta_t$ has been constrained by requiring that the accelerated electrons can produce synchrotron emission with spectral index $\sim1.1-1.5$ between 327 and 1400 MHz (Cassano \& Brunetti 2005). 
The spectral index of the radio halo in A2256 is $\sim 1.6$ and the relative velocity between merging sub-clusters (PC+SC) is $\sim2000$ kms$^{-1}$ (Berrington et al 2002). 
 The timescale for the decay of turbulence ($\tau_{kk}\sim$ Gyr) is comparable to the crossing time for merging subclusters. If the mergers in A2256 have occurred over the last Gyr then the presence of Mpc scale radio halo is consistent with the timescale over which the turbulence decays.
\par Further we compare the spectral trends and the geometries of mergers in A2256. 
The mean radial velocity of the Gr is $\sim2000$ kms$^{-1}$ higher than that of the PC and is moving along the line of sight into the PC (Miller et al 2003). Such a merger event could lead to a shock travelling from the SE to the NW and also inject turbulence in the swept ICM if the direction of merger is inclined to the line of sight. The shock and turbulence in the NW region and the turbulence in the region that was swept by the passage of the Gr can generate diffuse radio emission. Moreover the major merger between PC and SC must also result in injection of turbulence in the ICM. 
The spectral index trend in the diffuse radio emission(steepening from NW to SE)  is consistent with the geometry of the merger of the Gr with the PC. Miller et al (2003) point out the frequent occurrence of radio sources associated with the star forming galaxies in the Gr and interpret it as an effect of merger. Thus there is evidence for the Gr to have undergone merger in a direction into the plane of the sky; the exact orientation is not clear. Miller et al (2003) have argued that the merger of the Gr viewed 0.3 Gyr after the core passage is responsible for the diffuse radio emission in the NW of the cluster.
The co-spatial occurrence of the brightest diffuse radio emission and the Gr and the direction of spectral index steepening being consistent with the proposed direction of the Gr - PC merger, support the picture of Miller et al (2003). The radiative lifetimes discussed earlier also support the passage of the Gr within the last 0.4 Gyr. If the merger between PC and SC has been in progress for the past Gyr it will contribute to enhance the radio emission but its role in producing the spectral index distribution is not clear. Another possibility is that the amount of energy that was injected as fluid turbulence in different regions is different and thus has resulted in flat spectrum emission in one region and steep spectrum emission in another. The dependence of acceleration by turbulence on the many unknown parameters mentioned earlier make further progress in testing this possibility difficult.
\par The group Gr, being less massive and not being part of a major merger in A2256 is not favoured by CE06 to have created the radio emission in the NW region. Instead, CE06 have proposed two scenarios involving the major merger between the PC and the SC (see Fig. 11 a and 11 b in CE06); in both the scenarios the SC approaches the PC from the NW in projection on the plane of the sky. In the first scenario the SC approaches the PC from the NW and is proposed to be in the early stages of merger. In this picture the merger shock crosses from the NW to the SE or is along the line of sight towards the observer (see Fig. 11a in CE06). This implies steep spectra in the NW edge and flatter spectra towards SE or a uniform distribution of spectral indices as seen by the observer, respectively. The diffuse radio emission in the SE (radio halo) is considered a remnant of an older merger (CE06).  This picture is contradictory to the observed spectral steepening from the NW to the SE and does not account for the radio emission in the SE region (radio halo). In the second scenario of CE06, the SC is in an advanced stage of merging. The outgoing merger shocks are proposed to create the radio emission. One of the shock waves is along the line of sight towards the observer (Fig. 11b) and thus might result into complex spectral index distribution; it cannot explain the spectral index variation from the NW to the SE. 
\par Apart from the trend of spectral steepening from NW to SE,
the spectrum of the diffuse radio emission steepens at lower frequencies (Fig. 4). Since a synchrotron spectrum is expected to steepen at higher frequencies due to energy losses, 
this low frequency steepening in the diffuse emission cannot be explained by a single population of emitting particles. Superposition of at least two spectra having unequal amount of steepening can give rise to a spectrum that steepens at low frequencies.
It is possible that the two superposing spectra are due to populations of electrons accelerated at different epochs. The two merger events in A2256 could be the two epochs at which the electrons were accelerated; the merger of SC with PC and that of the Gr with the PC and the SC.
According to the models proposed to explain the X-ray substructure, the merger between the PC and the SC is viewed 0.2 Gyr prior to the core passage and the Gr merger is viewed 0.3 Gyr after the core passage (Miller et al. 2003; Roettiger et al 1995). These timescales are consistent with the timescale of $\sim0.08 - 0.4$ Gyr over which the radio emission remains detectable in the frequency range of 150 -1400 MHz. Due to lack of detailed simulations, the chronology between the two merger events and the exact mechanism of acceleration of electrons to relativistic energies cannot be established with confidence. Hydro/ N body simulations reproducing the optical and X-ray substructure and the temperature distribution are required to obtain the detailed geometries and the chronological order of the two mergers in A2256.
 An alternative possibility is that a complex distribution of turbulence and magnetic fields in the cluster may result in regions with flat and steep spectra. Such regions seen projected  along the line of sight could also produce a low frequency steepening in the integrated spectrum.
\par A peculiar low frequency steepening trend seen in the integrated spectra of A2256 and of the diffuse radio emission in it have been discussed above. It should be noted that at high frequencies, the synchrotron spectra are expected to steepen due to energy losses. The synchrotron emission due to acceleration mechanisms such as turbulence or shocks have a cutoff due to the maximum energy that is available in turbulence or the strength of the shock. In the case of A2256, a spectral index map of the NW region (marked G and H in the Fig. 2 of this paper and referred to as the relic by CE06) between 1369 and 1703 MHz (Fig. 4 in CE06) is available. This spectral index map in CE06 is affected by the loss of sensitivity to extended structure at 1703 MHz due to instrumental limitations and larger noise level. Nevertheless, the spectral indices in the high surface brightness relic region can be used for comparison with our spectral index maps. The spectral index map in CE06 shows a flat spectrum ($\alpha\sim-0.9$) emission at the NW edge and a steeper spectrum ($\alpha\sim-1.6$) emission in the SE region of the relic. 
The same relic region (marked G and H) in our spectral index map between 350 and 1369 MHz (Fig. 2, left) shows flatter spectral indices $\sim -0.7$ to $-1.0$. 
The steeper spectral indices in the 1369-1703 MHz spectral index map in CE06 could be the effect of ageing of the synchrotron spectrum due to energy losses. 
Thus putting together the spectral information across the frequency range of 150 to 1703 MHz for the NW region a picture consistent with a high frequency steepening due to the energy losses and a low frequency steepening due to a second component of electron population emerges. Further deeper observations at high  frequencies ($>1.4$ GHz) which can image the entire extent of the diffuse emission in A2256 are required to confirm the steepening due to energy losses. 
\section{Conclusions}
We have carried out a multi-wavelength analysis of the merging rich cluster of galaxies A2256. Using new radio observations at 150 MHz from the GMRT and archival observations from the VLA (1369 MHz) and the  WSRT (350 MHz), we have produced spectral index images of the diffuse radio emission in A2256 over the range 150-1369 MHz. These spectral index images show regions of the diffuse radio emission having flat spectral indices in the NW and having steep spectral indices in the SE of the cluster centre.
The implied synchrotron life times for the relativistic plasmas are in the range 0.08 - 0.4 Gyr. Such a distribution of spectral indices is interpreted as resulting from a merger through the cluster from the SE to the NW in the last 0.5 Gyr or so. Acceleration due to shocks can explain the emission only in the NW relic region. 
The generation of the diffuse radio halo emission is likely to be due to the turbulence injected in the ICM by the mergers. The injection of fluid turbulence by mergers in the last Gyr and the timescales of $\sim$Gyr required for the cascade of MHD turbulence are consistent with the spectral ages $\sim0.4$ Gyr.
Furthermore, the diffuse radio emission shows spectral steepening toward lower frequencies. This low frequency spectral steepening is consistent with a combination of spectra from two populations of relativistic electrons created at two epochs (two mergers) within the last $\sim$0.5 Gyr. Earlier interpretations of X-ray and optical data suggested that there have been two mergers in Abell 2256 in the last $\sim$0.5 Gyr, consistent with the current findings. 
Also highlighted is the futility of correlating the average temperatures of thermal gas with the average spectral indices of diffuse radio emission in respective clusters. 
Spectral index imaging of diffuse radio emission in galaxy clusters is found to be a powerful tool to study cluster mergers, dynamics, and history of cluster formation.
\begin{figure}
\epsscale{0.8}
\plotone{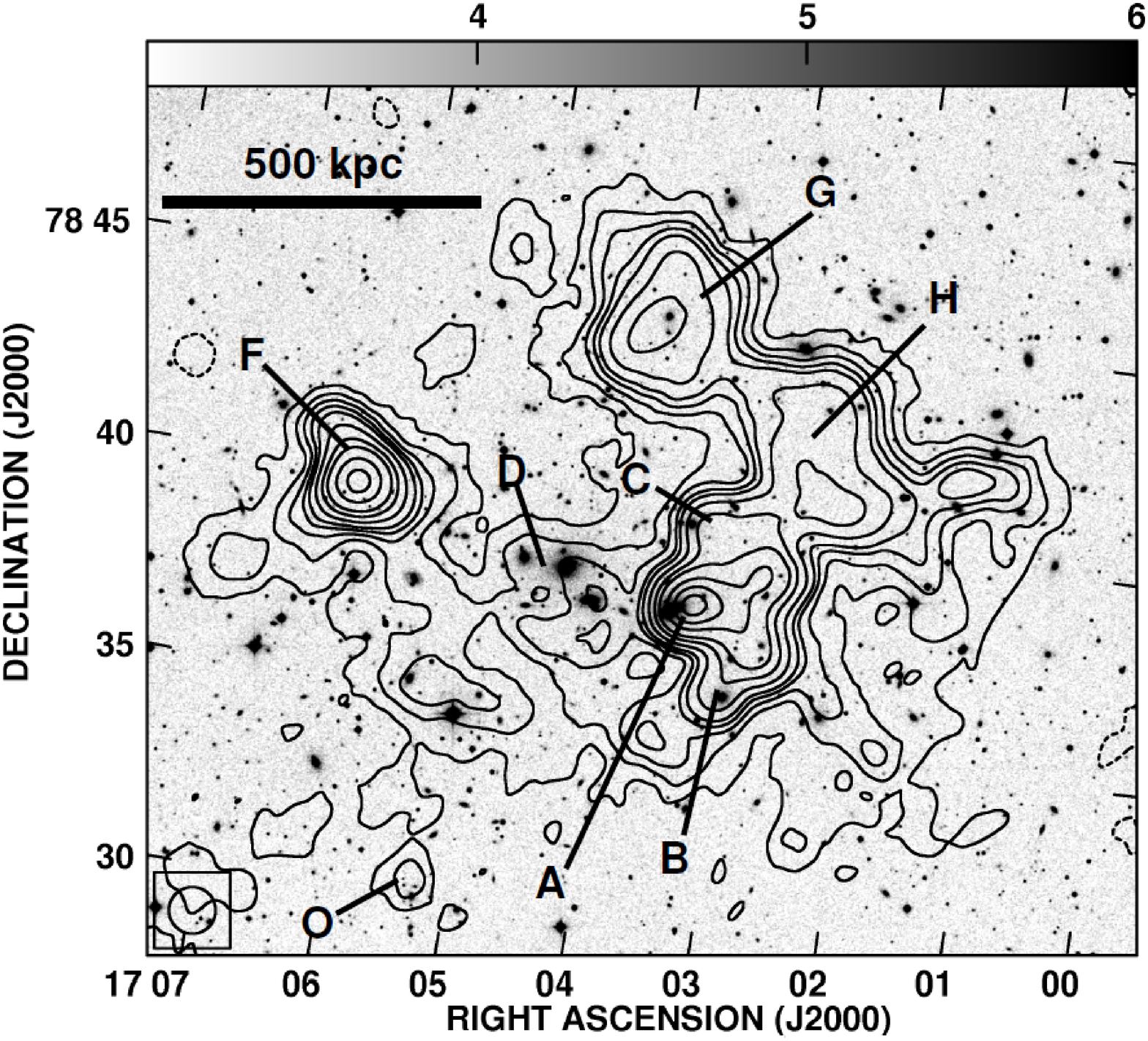}
\caption{A2256: GMRT 150 MHz image with a synthesised beam of $67''\times67''$ and an rms of $3.6$ mJy beam$^{-1}$ . Contours are at -18, 18, 36, 54, 72, 90, 126, 180, 240, 360, 480, 600 mJy beam$^{-1}$. The DSS R band image is shown in greyscale. The linear scale is marked for an assumed redshift of 0.058.
The labels A, B, C, D, F, O, G and H for the radio sources are according to the convention in Bridle et al (1979) and R94. \label{fig1}}
\end{figure}
\begin{figure}[h]
\epsscale{1}
\plotone{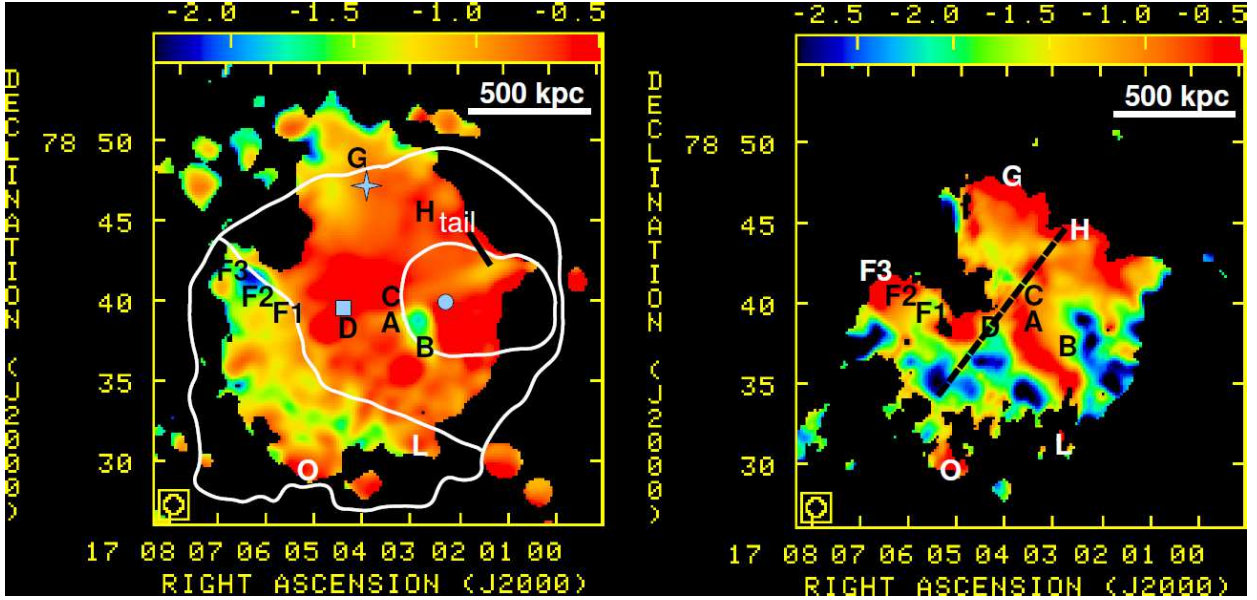}
\caption{Grayscale (colour in the online version) represents the spectral index of radio emission in A2256 between 350 and 1369 MHz({\it left}) and between 150 and 350 MHz ({\it right}). The synthesized beam is $67''\times67''$. The symbols square, circle and star (left panel) represent the centroids of the primary cluster, the sub-cluster and the group, respectively (Berrington et al 2002). In the left panel, the black (white in the online version) contours are a schematic representation of regions at different temperatures (Bourdin \& Mazzotta 2008). Contours enclose regions with temperatures of 4-6 keV (innermost), 6-8 keV (region surrounding the innermost region) and 8-10 keV (region toward SE) respectively. \label{fig2}}
\end{figure}
\begin{figure}[h]
\epsscale{1}
\plottwo{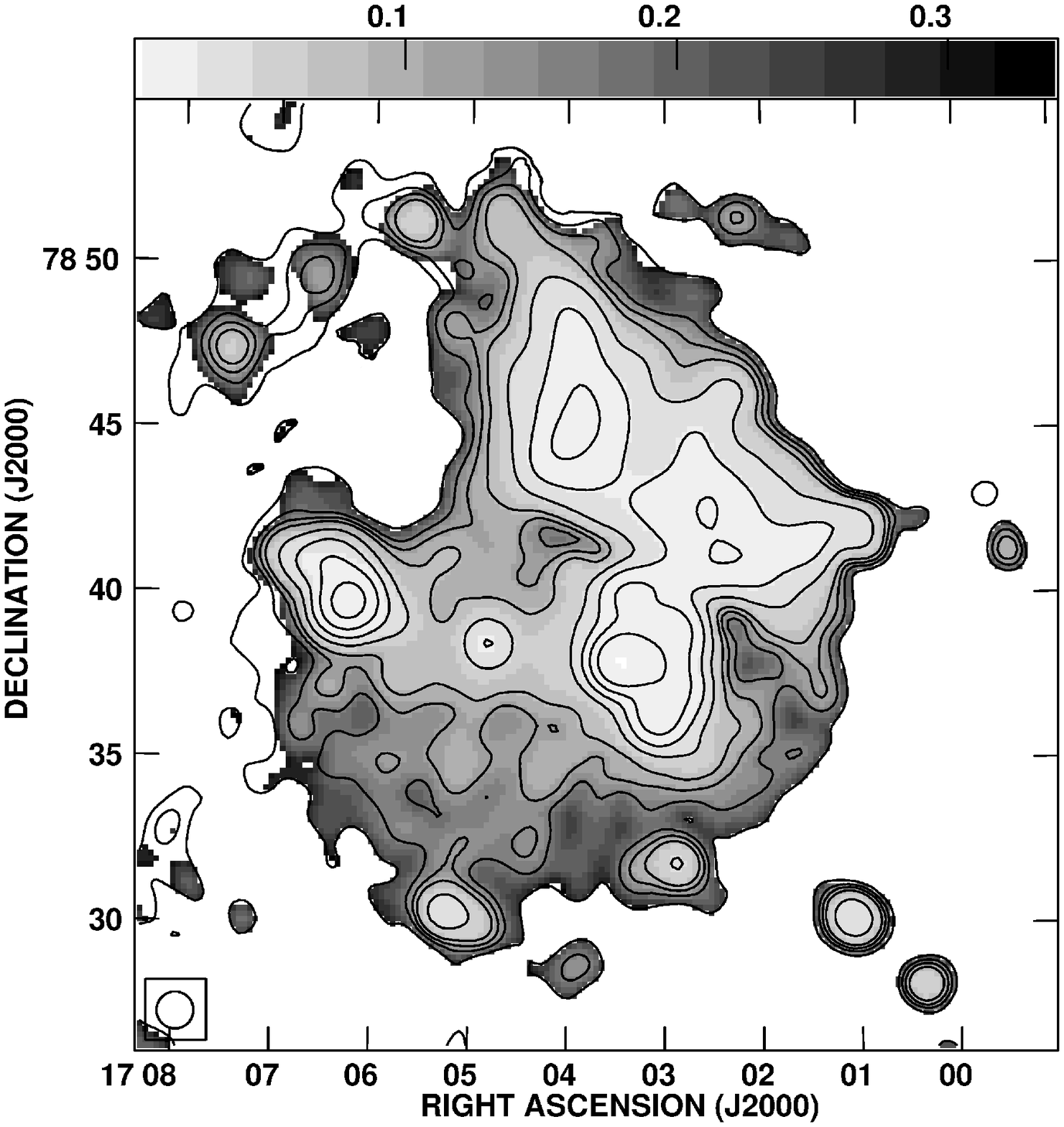}{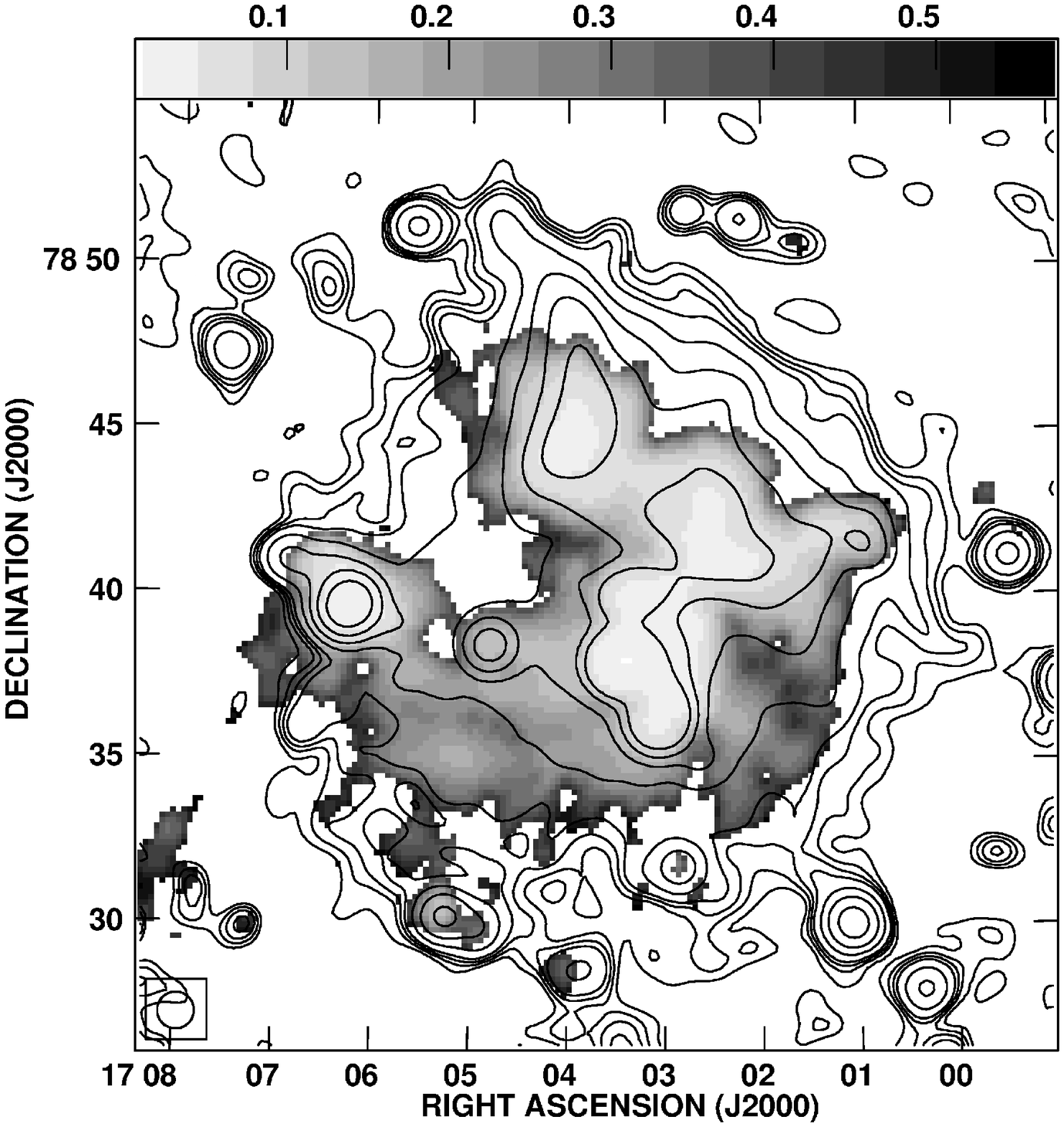}
\caption{Spectral index error maps in greyscale. {\it Left} 350-1369 MHz. {\it Right} 150-350 MHz. Overlaid are the images at 350 and 1369 MHz in the left and the right panels respectively. Synthesized beams in both the panels are $67''\times67''$. Contours (starting at $5\sigma$) are at: {\it (Left)} 3.0, 4.8, 6.6, 9.0, 18.0, 36.0, 72.0, 144.0 mJybeam$^{-1}$ and {\it (Right)} 0.35, 0.56, 0.77, 1.05, 2.1, 4.2, 8.4, 16.8 mJybeam$^{-1}$. \label{fig3}}
\end{figure}
\begin{figure}[h]
\epsscale{0.4}
\plotone{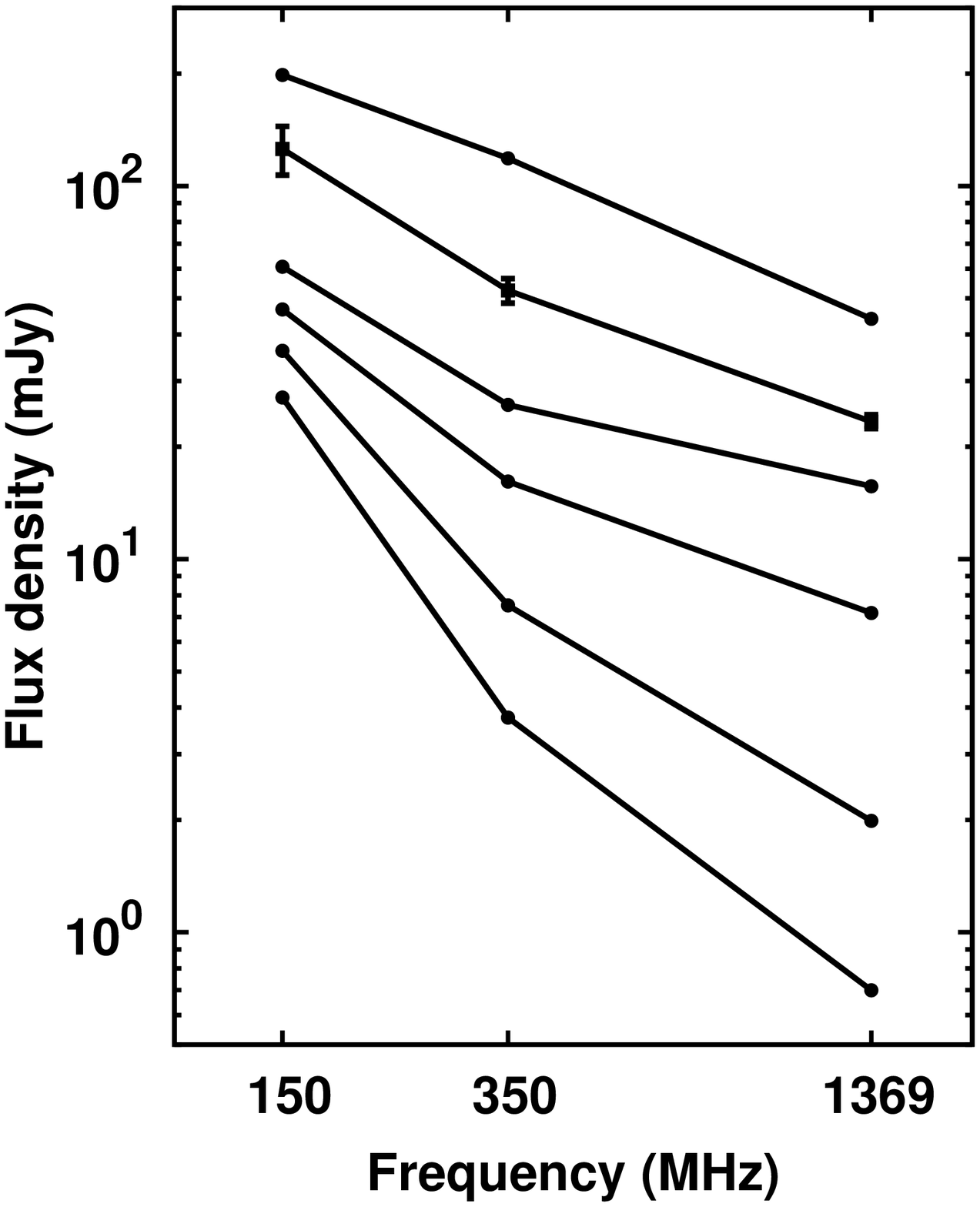}
\caption{Spectra of diffuse radio emission from six patches spaced equally along the broken line shown in
Fig. 2 (right panel). The spectral indices, ($\alpha_{150}^{350}$, $\alpha_{350}^{1369}$), for the spectra from top to bottom are: (-0.60, -0.72), (-1.03, -0.59), (-1.00, -0.36), (-1.25, -0.59), (-1.85, -0.97) and (-2.30, -1.23). Typical errors are indicated in the second spectrum from the top for which there is no scaling in the flux density. All other spectra are shifted along the y-axis. \label{fig4}}
\end{figure}
\acknowledgments
We thank the staff of the GMRT who have made these observations possible. GMRT is run by the National Centre for Radio Astrophysics of the Tata Institute of Fundamental Research. We thank the anonymous referee for valuable suggestions regarding shock acceleration and turbulence. This research has made use of the NASA/IPAC Extragalactic Database (NED) which is operated by the Jet Propulsion Laboratory, California Institute of Technology, under contract with the National Aeronautics and Space Administration. The National Radio Astronomy Observatory is a facility of the National Science Foundation operated under cooperative agreement by Associated Universities, Inc. The Westerbork Synthesis Radio Telescope is operated by the ASTRON (Netherlands Institute for Radio Astronomy) with support from the Netherlands Foundation for Scientific Research (NWO). \\
{\it Facilities:} \facility{GMRT (NCRA,TIFR)}, \facility{WSRT (ASTRON)}, \facility{VLA (NRAO)}.

\end{document}